\documentclass[12pt,english]{article}
\usepackage[T1]{fontenc}
\usepackage[latin9]{inputenc}
\usepackage{geometry}
\usepackage{verbatim}
\usepackage{amssymb}
\usepackage{setspace}
\usepackage{babel}
\usepackage{amssymb,amsmath}
\usepackage{graphicx, xcolor, varwidth}
\usepackage{hyperref}
\usepackage{caption}
\usepackage{subcaption}

\newcommand{\bea}{\begin{eqnarray}}
\newcommand{\eea}{\end{eqnarray}}
\newcommand{\be}{\begin{equation}}
\newcommand{\ee}{\end{equation}}

\usepackage{cite}

\numberwithin{equation}{section}



\begin{document}

\begin{titlepage}

\begin{center}
{\Large \textbf {Comments on Thermalization in 2D CFT}}

\vspace*{12mm}

Jan de Boer$^1$ and Dalit Engelhardt$^2$

\vspace*{8mm}

$^1$ \textit{Institute for Theoretical Physics, University of Amsterdam, 1090 GL Amsterdam, The Netherlands}\\
$^2$ \textit{Department of Physics and Astronomy, University of California, Los Angeles, CA 90095, USA}\\

\vspace*{8mm}

{\tt J.deBoer@uva.nl, engelhardt@physics.ucla.edu}

\vspace*{6mm}
\end{center}
\begin{abstract}

We revisit certain aspects of thermalization in 2D CFT. In particular, we consider similarities and differences between
the time dependence of correlation functions in various states in rational and non-rational CFTs.
We also consider the distinction between global and local thermalization and explain how states obtained by acting with a diffeomorphism on the ground state can appear locally thermal, and we review why the time-dependent expectation value of the energy-momentum tensor is generally a poor diagnostic of global thermalization. Since all 2D CFTs have an infinite set of commuting conserved charges, generic initial states might be expected to give rise to a generalized Gibbs ensemble rather than a pure thermal ensemble at late times. We construct the holographic dual of the generalized Gibbs ensemble and show that, to leading order, it is still described by a BTZ black hole. The extra conserved charges, while rendering $c < 1$ theories essentially integrable, therefore seem to have little effect on large-$c$ conformal field theories.

\end{abstract}

\end{titlepage}

\vskip 1cm

\tableofcontents

\section{Introduction}

Systems perturbed away from equilibrium have been shown to exhibit a rich array of behaviors that depend on the type of initial perturbation and the characteristics of the systems. At asymptotically late times, however, they are generically expected to exhibit behavior characteristic of thermal equilibrium, regardless of the short-time behavior following the perturbation, so long
as the perturbation injects sufficient energy into the system.
This behavior can be investigated from the point of view of a subsystem, where the system is defined to have thermalized if its reduced density matrix is equal to that of a thermal (mixed) state; however, isolated quantum systems starting from a pure quantum mechanical state can also be described as ``thermalized'' if the expectation values of observables at late times are in agreement with those of a thermal ensemble~\cite{Srednicki:1994aa,Deutsch:1991aa,Rigol:2008aa,Rigol:2012aa}. 

Theoretical and experimental evidence on non-equilibrium behavior
has increasingly shown that the expectation that all systems equilibrate
to a simple thermal state may be misleading in 1+1-dimensional systems, where
in some cases thermalization appears to be inhibited entirely\footnote{See~\cite{Daley,Eisert:2015aa,Langen:2016vdb} for recent reviews.}. In holographic contexts, where thermalization in the field
theory is understood to be dual to black hole formation in the holographically
dual asymptotically anti-de-Sitter (AdS) bulk spacetime, such non-thermalization
would appear to be at odds with the expectation that perturbations
of AdS of sufficiently high energy should generically backreact to form black holes\footnote{The question of whether exact, non-trivial, eternally oscillating asymptotically-AdS solutions (with energies equal to that of a large AdS black hole)
exist is currently an open issue; we comment on it briefly in Sec.~\ref{sec:discussion}.}.

Absence of thermalization in certain 1+1-dimensional isolated systems has been
attributed to the presence of a large number of conserved currents
and is associated with the notion of quantum integrability, with such systems proposed to behave in accordance with the generalized Gibbs ensemble (GGE) instead of the usual Gibbs ensemble~\cite{Rigol:2007aa}. The critical behavior of many 1+1-dimensional systems is described
by conformal field theories (CFTs), suggesting that there may be qualitative differences 
between the thermalization behavior of 2D CFTs with their infinite dimensional conformal
symmetry as compared to that of higher-dimensional CFTs, where the stress tensor and its descendants are the only
conserved currents to be found.

Nonetheless, even for unitary 2D CFTs there are important differences
between the behavior of CFTs whose central charge is below or above some critical
value $c_{\text{crit}}$, where $c_{\text{crit}}$ depends
on the chiral algebra of the 2D CFT and determines whether the CFT
is rational ($c<c_{\text{crit}}$) or not\footnote{While this appears to be the case in known examples, we are not aware
of a rigorous proof of this statement.} ($c>c_{\text{crit}}$).
For CFTs whose symmetry is only the Virasoro algebra, i.e. with no
additional extended symmetries, $c_{\text{crit}}=1$. In the rational
case, the spectrum of the theory consists of a finite number of primaries
for the chiral algebra with rational conformal dimensions of the form
$\frac{p}{q}$, with integer $p,q$. For a CFT on a circle of radius
$R$, time translations are thus generated by $U(t)=\exp\left(- i t\left(L_{0}+\bar{L}_{0}-\frac{c}{12}\right)/R\right)$,
so that all correlation functions will be periodic in time\footnote{It should be emphasized that these revivals are different from Poincar\'{e} recurrences and occur on far shorter time scales.}, where the existence of such revivals follows from the rationality of the conformal dimensions in the CFT, and their period depends on the operator spectrum and on the size $R$ of the
system. Clearly, theories with $c<c_{\text{crit}}$ do not thermalize, although
it is still in principle possible for subsystems to behave approximately
as thermal systems for times $t$ that are much smaller than the revival
time of the system.

For $c>c_{\text{crit}}$, this argument no longer applies, and there
is no \textit{a priori} mechanism to prevent thermalization for generic
perturbations. There are nonetheless special states that fail to thermalize
in any CFT, the simplest examples being states that are built from
descendants of the ground state only. These states are linear superpositions
of states with integer conformal dimension,
and their period is proportional to the system size $L$ alone. Even in such states, sufficiently
small subsystems will exhibit approximately thermal behavior for times
$t\ll L$, however globally the system undergoes periodic
revivals.

Special descendants of the ground states, coherent states, have a geometric interpretation as
conformal transformations of the CFT on the plane or a subspace thereof.
In the case of a bounded subspace, such boundary states in the form
of a strip or a rectangle have been used to analyze certain quantum
quenches and CFT non-equilibrium behavior (see, e.g., \cite{Calabrese:2006rx,Calabrese:2007rg,Calabrese:2007aa,Hartman:2013qma,Cardy:2014rqa,Kuns:2014zka,Engelhardt:2015axa,Asplund:2015eha,Calabrese:2016xau}),
with conditions on the Euclidean boundary defining the initial conditions
of the system, whose time-evolved correlation functions are computed
by analytic continuation from Euclidean time. Such states can be understood to define an initial state via a Euclidean path integral over a portion of the boundary, with correlation functions computed by joining together domains representing an in- and an out-state. 
For example, a path integral over over a rectangle with suitable boundary conditions on three sides provides a state in the CFT on the interval formed by the remaining side; a correlation function in this state can be computed by joining together such an in-state with an out-state, resulting in a full rectangle\footnote{This is discussed in more detail in Sec.~\ref{sec:bndrystateformalism}.}. Similarly, in the case of the strip opposite halves represent the in- and out-states. 
While strip states have
been shown to exhibit behavior consistent with thermalization, one has to be
careful with CFTs defined on the entire real line. Paraphrasing the result of 
\cite{Kuns:2014zka}, a conformal compactification of the real line maps it to
a finite interval\footnote{The precise map is $e^{i\tilde{z}}=(e^z-1)/(e^z+1)$,
where $z\in [-\infty,\infty]\times[0,\pi]$ is the coordinate on the Euclidean strip
of infinite spatial extent, and $\tilde{z}\in [0,\pi]\times[-\infty,\infty]$ a coordinate on a finite strip
with infinite extent in Euclidean time.}, and it maps all of Minkowski spacetime to a causal diamond
based on the finite interval. Therefore, measurements in Minkowski space are insensitive to the presence of the boundaries
of the interval, which introduce a finite size in the system that
determines the period of revivals. The restriction of strip observations
to a causal subset thus prevents non-thermal features of such states
from being detected. This paper therefore made it clear that the apparent
thermalization in strip states observed in \cite{Calabrese:2006rx,Calabrese:2007rg,Calabrese:2007aa,Hartman:2013qma}
is due to the restriction to a limited amount of time. 

These observations can be further motivated by noting that many features of
global thermalization of a CFT, such as the appearance
of a suitable coarse-grained entropy, should be conformally invariant.
In fact, one could argue that a better (and conformally invariant) definition of thermalization would be
to require that expectation values at late times approach those of a thermal
state or those of a conformally transformed thermal state.
In particular,
in holographic theories, where conformal mappings are dual to
bulk diffeomorphisms, thermalization invariance under conformal mappings
is equivalent to the evident statement that black hole formation (or lack
thereof) is diffeomorphism invariant, and that asymptotic (AdS boundary)
observers must agree on whether or not a black hole has formed in
the bulk. Since black hole formation following an injection of energy
is rather generic in AdS,
this calls into question which CFT states do in fact thermalize. As
we show here, in non-rational CFTs, i.e. where $c>c_{\text{crit}}$, no revivals would be observed in expectation values of primary operators in general states constructed as linear superpositions of states obtained by local operator insertions. To the extent that the absence of revivals in the system is indicative of its thermalization, this is in line
with the expectation from holography.

An interesting additional feature of 2D CFTs is the existence of an infinite number of commuting conserved charges, even when the chiral algebra is just the Virasoro algebra. The lowest two charges are $L_0$, the zero mode of $T$, and the zero mode $K_0$ of $:TT:$. These charges are a quantum version of the infinite number of conserved charges that appear in the KdV hierarchy~\cite{Bazhanov:1994ft}. One would more generally expect that generic states in a 2D CFT at late times should be describable in terms of a generalized Gibbs ensemble with chemical potentials for all conserved charges instead of the thermal ensemble. This has indeed been confirmed in \cite{Cardy:2015xaa,Mandal:2015jla,Mandal:2015kxi}. 

A nice heuristic picture of some of the features of thermalization
in 2D CFTs arises by assuming that all excitations can be described
in terms of free quasi-particles~\cite{Calabrese:2005in,Calabrese:2006rx,Calabrese:2007rg,Calabrese:2009qy}.
If after a quench correlated pairs of quasi-particles are locally
emitted, the entanglement between an interval of length $L$ and its
complement will increase until time $T\sim L/2$ and then remain constant. This picture of growth and saturation is qualitatively in keeping with the holographic predictions \cite{Balasubramanian:2010ce,Balasubramanian:2011ur,Caputa:2013eka,Liu:2013qca}. In the case of a union of disjoint intervals, on the other hand, the post-quench behavior of the entanglement entropy given by the quasi-particle picture only correctly corresponds to the behavior for $c<c_{\text{crit}}$ \cite{Asplund:2015eha} systems. There therefore
appear to be close connections between integrability, rational conformal
dimensions, and the validity of the quasi-particle picture for $c<c_{\text{crit}}$
on the one hand, and between irrational conformal dimensions, lack
of integrability, and the breakdown of the quasi-particle picture
for $c>c_{\text{crit}}$.

The inhibition of thermalization that we find in rational CFTs by
contrast to general CFTs thus further asserts such connections. In
this paper, we clarify some additional aspects of these connections
and make contact with the dual holographic picture that they provide.
We begin by discussing the holographic dual picture of local thermalization
in a pure state and analyze the capacity of the CFT stress tensor
for serving as a thermalization diagnostic (Sec.~\ref{sec:thermprobes}).
We then exploit the conformal invariance of global thermalization
in a CFT by evaluating whether local perturbations of the rectangle
state are followed by initial-value revivals of observables at asymptotically-late
times; such revivals are indicative of the system's inability to establish
an asymptotic thermal state, and we show that as is expected from holography, they naively
do not take place for a general
(non-rational) CFT (Sec.~\ref{sec:spectrumdependence}). This discussion
is preceded by a review of the boundary-state setup and the strip
and rectangle states (Sec.~\ref{sec:bndrystateformalism}). Finally,
we consider the holographic dual of the generalized Gibbs ensemble
with chemical potentials for all conserved charges and show that it
is still described by the BTZ black hole (Sec.~\ref{sec:GGE}). We
conclude with a discussion of future directions.

\section{Probes of local and global thermalization}\label{sec:thermprobes}

The general thermalization setup is to consider a CFT in a pure state $|\psi\rangle$, let the system time evolve, 
and ask to what extent the state of system can be well approximated by a thermal state (global thermalization) and 
to what extent a subsystem can be well approximated by a subsystem of a thermal system (local thermalization).

A unique feature of 2D CFTs is that they have an infinite symmetry algebra that creates new states
$|\psi'\rangle \sim\sum\prod L_{-k_i}|\psi\rangle$ from $|\psi\rangle$. We would expect that these symmetries
do not affect whether or not a system globally thermalizes, but it is not \textit{a priori} clear in what way these
symmetry generators affect local thermalization. The example of the rectangle state (which is related to 
the ground state by symmetries) shows that local thermalization can occur even in states that are descendants
of the ground state: by restricting observations to a small interval on the rectangle, the geometry observed is effectively that of the infinite strip and therefore thermalization is observed. It would be quite interesting to develop a more quantitative theory explaining to what extent subsystems in states that are descendants of the ground state are approximately thermal. Given that
the behavior of the systems of interest seems to be fixed by geometry and symmetries alone, such a quantitative description should be possible,
and we hope to report on it elsewhere. In the meantime we will present the holographic dual point of view.

In holography, states that are descendants of the ground state and that have a semiclassical gravitational dual
are described by geometries that are diffeomorphic to global AdS${}_3$. General descendants of the ground state
are described by AdS${}_3$ with many graviton excitations, and different semiclassical AdS${}_3$ geometries correspond to various Virasoro
coherent states. Diffeomorphisms that preserve a convenient Fefferman-Graham gauge choice act on AdS${}_3$ 
as follows. We start with vacuum AdS with metric $ds^2=(dw^2+dz d\bar{z})/w^2$, and perform the following coordinate transformation
\begin{equation}
w\rightarrow \frac{w\sqrt{\partial f \bar{\partial} \bar{f}}}{N},\quad
z\rightarrow f(z) -\frac{w^2}{2} \frac{\partial f \bar{\partial}^2 \bar{f}}{\bar{\partial}\bar{f} N},\quad
\bar{z} \rightarrow \bar{f}(\bar{z}) -\frac{w^2}{2} \frac{\bar{\partial} \bar{f} \partial^2 f}{\partial f N},
\end{equation}
where 
\begin{equation}
N=1+\frac{w^2}{4}\frac{ \partial^2 f \bar{\partial}^2 \bar{f}}{ \partial f \bar{\partial}\bar{f} } .
\end{equation}
We then obtain a metric of the form
\begin{equation} \label{genmet2}
ds^2 = \frac{dw^2 + dzd\bar{z}}{w^2} -\frac{6}{c}T(z) dz^2 -\frac{6}{c}\bar{T}(\bar{z})d\bar{z}^2 +\frac{36}{c^2} w^2 T(z) \bar{T}(\bar{z})dz d\bar{z}
\end{equation}
where 
\begin{equation}
 T(z)=\frac{c}{12} \{f,z\},   \quad
\bar{T}(\bar{z}) = \frac{c}{12}\{ \bar{f},\bar{z} \},
\label{eq:TandTbar}
\end{equation}
and the Schwarzian derivative is as usual
\begin{equation}
\{ f,z \} = \frac{\partial^3 f}{\partial f} - \frac{3}{2} \left( \frac{\partial^2 f}{\partial f} \right)^2 .
\label{eq:schwarzian}
\end{equation}

If we restrict to an interval where $T(z)$ and $\bar{T}(\bar{z})$ are approximately constant, then the 
bulk geometry in the neighborhood of that interval will be close to the BTZ geometry\footnote{The metric \ref{genmet2} corresponds to the metric of \cite{Banados:1998gg}
under the variable change $\rho=-\ln w$ and upon setting $c=\frac{3\ell}{2G}$.}~\cite{Banados:1998gg}, and correlation functions
computed there are approximately the same as the finite temperature correlation functions obtained from the BTZ
geometry. Thus in order to obtain local thermalization we should apply a diffeomorphism that produces a locally
constant $T(z)$ and $\bar{T}(\bar{z})$. An example of such a diffeomorphism is one that is locally approximately an
exponential $f(z) =\exp(\alpha z)$ as this has a constant Schwarzian derivative. This is not too surprising as
an exponential map essentially produces a local version of the Unruh effect whereby accelerated observers observe 
a thermal state. 

Globally, then, these diffeomorphisms produce what would appear as a local concentration of energy-density repeatedly oscillating (due to the global periodic time-dependence) in AdS, but that does not form a black hole even at arbitrarily late times. It is therefore clear that diffeomorphisms alone, absent additional energy injections into AdS, never produce global thermalization. This leads to 
an interesting reverse question: given the expectation values of $T(z)$ and $\bar{T}(\bar{z})$ in some state,
is it possible to come up with a diagnostic for whether or not the dual description of this state involves a black hole? In order to find such a diagnostic, we need to make sure that our diagnostic is not sensitive to diffeomorphisms, 
as the question of whether or not there is a black hole is clearly diffeomorphism invariant.

Perhaps the simplest way to analyze this problem is to find a diffeomorphism that makes $T(z)$ and $\bar{T}(\bar{z})$ 
constant and to read off the relevant constant values\footnote{There is a subtlety here, as such a diffeomorphism may not always exist. As nicely reviewed in \cite{Balog:1997zz}, one can classify the $T(z)$ that are inequivalent under diffeomorphisms of the circle,
which is the same as the classification of the so-called Virasoro coadjoint orbits. Besides the orbits which contain a point with constant $T(z)$, there are several other
orbits, but all of these orbits except one have an energy $L_0$ which is unbounded from below and are therefore most 
likely unphysical. The one remaining orbit, labeled ${\cal P}_1^-$ in \cite{Balog:1997zz}, has energy bounded from below,
and its physical relevance (if any) is not clear to us. In any case, if we use the Chern-Simons description, and use $SL(2,\mathbb R)$
gauge transformations instead of diffeomorphisms, we can always achieve constant $T$. We will ignore this subtlety in the remainder
of the paper and would like to thank Glenn Barnich for drawing our attention to this issue.}. 
If both are larger than $0$ in the planar case (or larger
than $c/24$ in the global case) then the dual description can possibly involve a black hole, whereas for smaller values
this is impossible and the system does not exhibit global thermalization. Note that this is a necessary, not a 
sufficient, condition for the existence of a black hole, as a large amount of dilute matter could also produce
the relevant energy densities without there being a black hole.

The Chern-Simons description of three-dimensional gravity suggests a different way to do this computation. Diffeomorophisms
act as gauge transformations on the SL$(2,\mathbb R)$ gauge field
\be 
A=\left( \begin{array}{cc} 0 & 1 \\ \frac{6}{c} T & 0 \end{array} \right)
\ee
and therefore the relevant constant values of $T$ can also be read off from the Wilson loop \cite{Martinec:1998wm}
\be \label{wilsonloop}
\cosh \frac{6}{c} T_{\rm const} = \frac{1}{2} {\rm Tr} P \exp \oint A dx.
\ee
One can think of the coordinates that yield constant values for $T(z)$ as the AdS${}_3$ analogue of the ``center of mass'' 
frame. 

As a side remark, the geometries (\ref{genmet2}) have recently been used to study gravitational hair for black holes,
see e.g. \cite{Donnelly:2015taa,Compere:2015knw,Sheikh-Jabbari:2016unm}, with $T(z)$ and $\bar{T}(\bar{z})$ playing
the role of the gravitational hair. From the Chern-Simons point of view
the only gauge-invariant observables in the theory are the Wilson loops (\ref{wilsonloop}), which commute with all the Virasoro generators
and which can be viewed as a Casimir for the Virasoro generators. These measure the invariant mass and angular momentum of the black
hole. By contrast, there is no gauge-invariant observable in Chern-Simons
theory that measures the gravitational hair away from the boundary of AdS or near the horizon of the black hole. In particular,
there is no observable in the interior of AdS in Chern-Simons theory that would allow one to detect the gravitational hair, 
suggesting that the hair has nothing to do with the degrees of freedom making up the black hole\footnote{We would like to thank
the participants of the Workshop on Topics in Three Dimensional Gravity (ICTP, Trieste) for useful discussions of these points.}. 

The above considerations are meant to illustrate that while the stress tensor alone may provide some indication of thermalization, it is not a sufficiently sensitive diagnostic. 
This can be further motivated by observing that in theories with holographic duals the stress tensor only captures the 
behavior of the metric near the boundary of AdS. The analysis of physics deep inside the bulk, including whether or not a black hole is
present, in general requires a knowledge of the expectation values of other operators in the theory as well. 

More generally, in arbitrary CFTs the expectation values of all the higher conserved charges can be rendered constant by acting with more complicated Virasoro symmetries (beyond diffeomorphisms). However, these higher conserved charges do not appear to play an important role in AdS/CFT, which we shall see for the case of 2+1 dimensions in Sec.~\ref{sec:GGE}.

Finally, we note that the holographic bulk geometries obtained via (\ref{genmet2}) are dual to conformal transformations of the CFT on the full plane. In order to apply this approach to find the holographic dual of arbitrary bounded subsets of this CFT, i.e. BCFTs, it is necessary to equip this description with an appropriately-chosen extension of the boundary of the CFT to the bulk - a bulk brane that bounds the spacetime region dual to this BCFT in the spirit of the AdS/BCFT correspondence of \cite{Takayanagi:2011zk,Fujita:2011fp}. Applying (\ref{genmet2}) to such setups in order to describe holographically a large class of holographic duals to BCFTs is an interesting direction that we leave to future work. Importantly, however, the presence of such a bulk brane is not expected to affect the local bulk physics in the deep interior (far away from the brane) of this spacetime, so that the above statements regarding local thermalization should carry over in the BCFT regime as well so long as the subsystem considered is sufficiently far from the boundary endpoints of the CFT.

\section{Non-equilibrium behavior from CFT boundary states\label{sec:bndrystateformalism}}

The setup underlying the CFT non-equilibrium dynamics approaches of \cite{Calabrese:2006rx,Calabrese:2007rg,Calabrese:2007aa,Hartman:2013qma,Cardy:2014rqa,Kuns:2014zka,Engelhardt:2015axa,Asplund:2015eha} -- and which lends a physical interpretation to the strip and rectangle states -- is that of the Calabrese and Cardy (CC) boundary state model for non-equilibrium evolution in CFTs ~\cite{Calabrese:2006rx,Calabrese:2007rg}. This boundary state setup relies on the existence
of a well-defined analytic continuation from Lorentzian to Euclidean
time in the system. This allows an initial state of the system $\left|\psi_{0}\right\rangle $
to be described as a Euclidean boundary state $\left|B\right\rangle $.
The system is taken to have a Hamiltonian $H$, and the initial
state $\left|\psi_{0}\right\rangle $ is assumed to be an eigenstate
of a different Hamiltonian $H_{0}$. Conformal boundary states are in fact non-normalizable, and in practice the quench is taken to be from a gapped Hamiltonian, so that the actual Euclidean boundary state is given by a state that is irrelevantly perturbed from the conformal boundary state $\left|B\right\rangle $; by convention it is taken to be
\begin{equation}
\left|\psi_{0}\right\rangle _{E}\propto e^{-\tau_{0}H}\left|B\right\rangle, 
\label{eq:bndrystateform}
\end{equation}
where $\tau_{0}$ is on the order of the correlation length of the gapped Hamiltonian $H_0$. We note that $H\propto\int T_{tt}dx$, where $T_{tt}\propto T(z)+\bar{T}(\bar{z})$, and in general additional irrelevant operators are expected to contribute. More general forms of boundary states where additional conserved charges or boundary operators are introduced in the exponential and act on the conformal boundary state were considered in~\cite{Cardy:2015xaa,Mandal:2015jla,Mandal:2015kxi}. The restriction to $T_{tt}$ in (\ref{eq:bndrystateform}) was motivated in \cite{Cardy:2014rqa} by noting that $T_{tt}$ is often the leading irrelevant operator acting on the
boundary state, and here we restrict our analysis to this form.

At $t=0$ the system is put in the state $\left|\psi_{0}\right\rangle $, and it is thereafter allowed to evolve unitarily
as $e^{-iHt}\left|\psi_{0}\right\rangle $. Correlation functions
of observables $\mathcal{O}(t,x)$ are therefore given by
\[
\left\langle \mathcal{O}(t,x)\right\rangle =\left\langle \psi_{0}\left|e^{iHt}\mathcal{O}(x)e^{-iHt}\right|\psi_{0}\right\rangle.
\]
and upon analytic continuation to Euclidean time can be computed via a path integral over a strip, of width $2\tau_0$, with the operator $\mathcal{O}$ inserted
at $\tau=\tau_{0}$ and analytically continued as $\tau\rightarrow\tau_{0}+it$.

In a 2D CFT, where the strip of width $2\tau_{0}$ can be conformally mapped to the upper-half plane
(UHP) as $w\rightarrow z(w)=e^{\frac{\pi}{2\tau_0}w}$, correlation functions
in this setup can simply be computed by conformal transformations
from the correlation functions of a boundary CFT (BCFT) on the UHP. This setup was used by CC to show that one-point functions
decay exponentially for $t\gg\tau_{0}$ and to compute the time evolution
of correlations between two primary operators (via the two-point function).

Since the restriction of the CFT to the UHP reduces the symmetry group
of the CFT, boundary conditions must be enforced at the interface
such that the conformal symmetry group is retained under conformal
maps from the UHP. These are given by the condition that there should
be no energy or momentum flow across the boundary, $\left.T_{xy}\right|_{y=0}=0$,
or 
\begin{equation}
\left.T(z)=\bar{T}(\bar{z})\right|_{z=\bar{z}}.\label{eq:conformalBC}
\end{equation}
In the presence of additional symmetries in the CFT, boundary conditions
that retain these symmetries may be imposed; however, the specification
of the BCFT alone does not require the boundary to respect these additional
symmetries. 

The implication of this conformal boundary condition is that the holomorphic
and anti-holomorphic sectors of the CFT are no longer independent.
In particular, $n$-point bulk correlators $\left\langle \phi_{h_{1},\bar{h}_{1}}\left(z_{1},\bar{z}_{1}\right)\phi_{h_{2},\bar{h}_{2}}\left(z_{2},\bar{z}_{2}\right)...\phi_{h_{n},\bar{h}_{n}}\left(z_{n},\bar{z}_{n}\right)\right\rangle$ on the upper-half plane obey the same Ward identities as the formal 
$2n$-point correlators of holomorphic
fields on the full plane~\cite{francesco1999conformal}, $$\left\langle \phi_{h_{1}}\left(z_{1}\right)\phi_{\bar{h}_{1}}\left(z_{1}^{*}\right)\phi_{h_{2}}\left(z_{2}\right)\phi_{\bar{h}_{2}}\left(z_{2}^{*}\right)...\phi_{h_{n}}\left(z_{n}\right)\phi_{\bar{h}_{n}}\left(z_{n}^{*}\right)\right\rangle.$$ 
The presence of the boundary thus implies that, e.g., one-point functions
of primary operators no longer vanish in general on the UHP and are
determined by conformal invariance up to a constant to have the form
$\left\langle \phi_{h,\bar{h}}(z,\bar{z})\right\rangle \sim\left(z-\bar{z}\right)^{-2h}$
for $h=\bar{h}$.

The infinite conformal symmetry of 2D CFTs allows a boundary state
defined on the upper-half plane to be mapped to an effectively unlimited
range of bounded domains, with the strip only one particular example;
as noted earlier, such mappings do not affect the thermal behavior
of the system, and whether or not the system reaches a global thermal state
is invariant under these transformations. Consequently, the out-of-equilibrium
behavior of a system from a particular boundary state can be investigated
in any of the conformally-equivalent boundary states. In the absence
of additional operator insertions, these states are simple conformal
mappings of the ground-state on the UHP and do not exhibit global thermalization.
As noted earlier, any of these boundary states can be mapped to the strip geometry, where the expectation value of the stress tensor is a constant Casimir value due to the vanishing of $\left\langle T(z)\right\rangle$ on the ground-state on the UHP.
The simplest modifications of the boundary states that potentially exhibit global thermalization are those obtained by
local operator insertions. As we show below, to diagnose thermalization of such systems, it is necessary to consider more refined observables. If additional operators are inserted on the boundary of the domain, conformal mappings do not affect the nature of these fields as boundary fields, since the boundary of a given system (e.g.
the $x$-axis on the UHP) is mapped to the boundary of the conformally
transformed system.

\subsection{Revivals in finite-length systems}

The finite-length equivalent of the CC setup is a boundary state defined on
a strip with spatial boundaries. Such bounded domains, with a vast
array of differently-shaped boundaries, can be obtained by Schwarz-Christoffel
maps~\cite{driscoll2002schwarz} from the UHP to bounded polygonal
geometries. These transformations map a set of designated
prevertices on the real line of the complex plane to the vertices
of a new polygonal domain, with the real line mapped to the boundary
of the domain. In particular, we can consider the map to a rectangle.
\begin{figure}[h]
\begin{center}
  \includegraphics[width=1.0\textwidth]{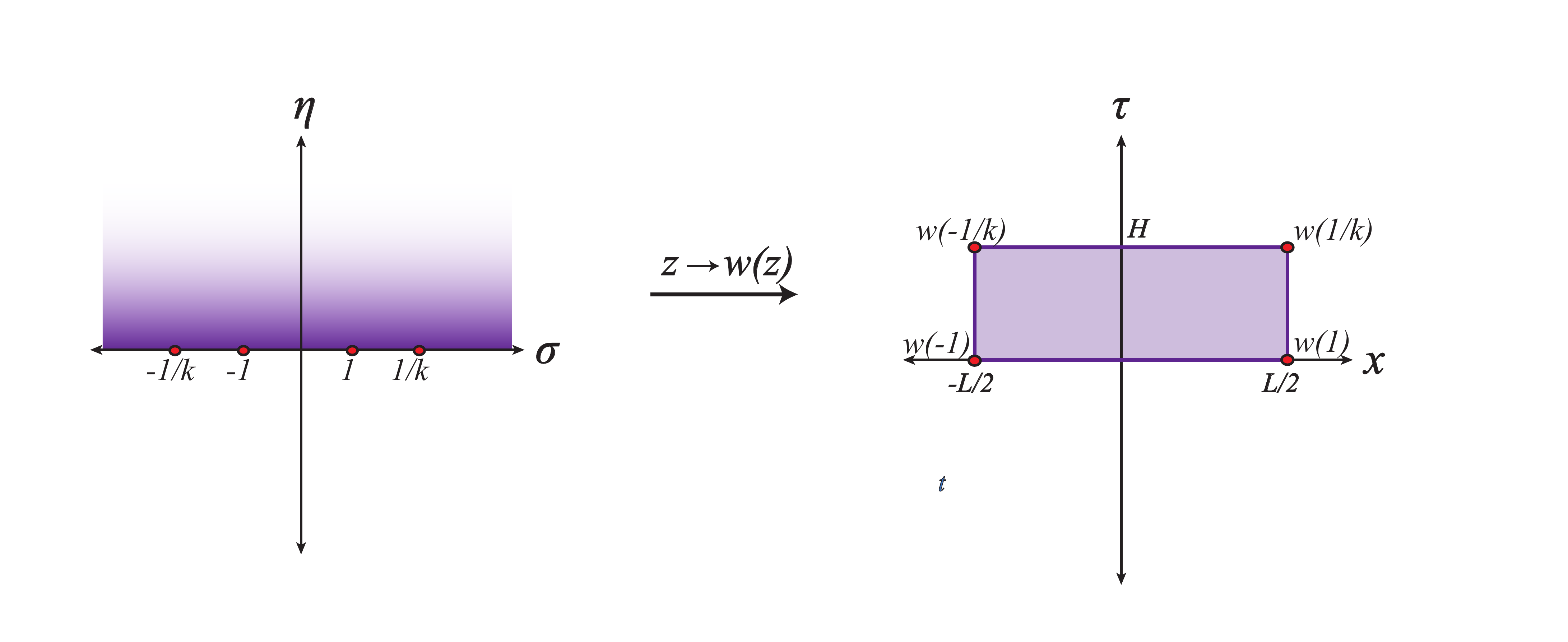}
\end{center}
\vspace{-0.3in}
\caption{\small{The Schwarz-Christoffel conformal transformation that maps the UHP with designated prevertices to a rectangle.}}
  \label{fig:UHPtoRec}
\end{figure}
For prevertices at
$x=\pm1,\pm\frac{1}{k}$, the general form of the map $z\rightarrow f(z)=w$
to the rectangle is given by the integral expression
\[
w(z)=A\int_{0}^{z}\frac{d\zeta}{\left(\zeta-1\right)^{\frac{1}{2}}\left(\zeta+1\right)^{\frac{1}{2}}\left(\zeta-\frac{1}{k}\right)^{\frac{1}{2}}\left(\zeta+\frac{1}{k}\right)^{\frac{1}{2}}}
\]
where $A$ is a constant that can be freely chosen. With a choice
of $A=-\frac{L}{2kK_{1}\left(k^{2}\right)}$, where $K_{1}\left(k^{2}\right)$
is the complete elliptic integral of the first kind and $k\in\left[0,1\right]$,
$w(z)$ is given as an elliptic integral of the first kind
\begin{equation}
z\rightarrow w(z)=\frac{L}{2K_{1}\left(k^{2}\right)}F\left(\arcsin z,k^{2}\right)
\label{eq:UHPtoRec}
\end{equation}
and maps the UHP to a rectangle with vertices at $\left(\pm\frac{L}{2},0\right)$
and $\left(\pm\frac{L}{2},H\right)$, where $H=\frac{K_{1}\left(1-k^{2}\right)}{2K_{1}\left(k^{2}\right)}$
is the height of the rectangle (Fig.~\ref{fig:UHPtoRec}).
The geometry of the rectangle is fully determined by the ratio $L/H$.
The limit of $k\rightarrow1$ corresponds to the zero-height rectangle,
and in this limit the system appears infinite in length. The limit of $k\rightarrow0$ corresponds
to the semi-infinite strip with width $L$. 

The inverse map from the rectangle to the UHP is given by the elliptic
Jacobi function 
\begin{equation}
w\rightarrow z(w)=\text{sn}\left(\frac{2K_{1}\left(k^{2}\right)}{L}w,k^{2}\right),\label{eq:ellipticJacobi}
\end{equation}
which is periodic in its argument as
\[
\text{sn}\left(\frac{2K_{1}\left(k^{2}\right)}{L}\left(w+mL+2inHL\right),k^{2}\right)=\left(-1\right)^{m}\text{sn}\left(\frac{2K_{1}\left(k^{2}\right)}{L}w,k^{2}\right).
\]
We will denote the complex coordinate on the rectangle by $w=x+i\tau$,
with $\tau$ the Euclidean time direction along
the height of the rectangle and $x$ the direction along its width. Observables
in this geometry are inserted on the rectangle
and analytically continued to Lorentzian times as $\tau\rightarrow \frac{H}{2}+it$,
where $t$ denotes the Lorentzian time coordinate. As a result, (\ref{eq:ellipticJacobi})
is periodic in Lorentzian time with period equal to $2L$.
\begin{figure}
    \centering
    \begin{subfigure}[b]{0.57\textwidth}
        \includegraphics[width=\textwidth]{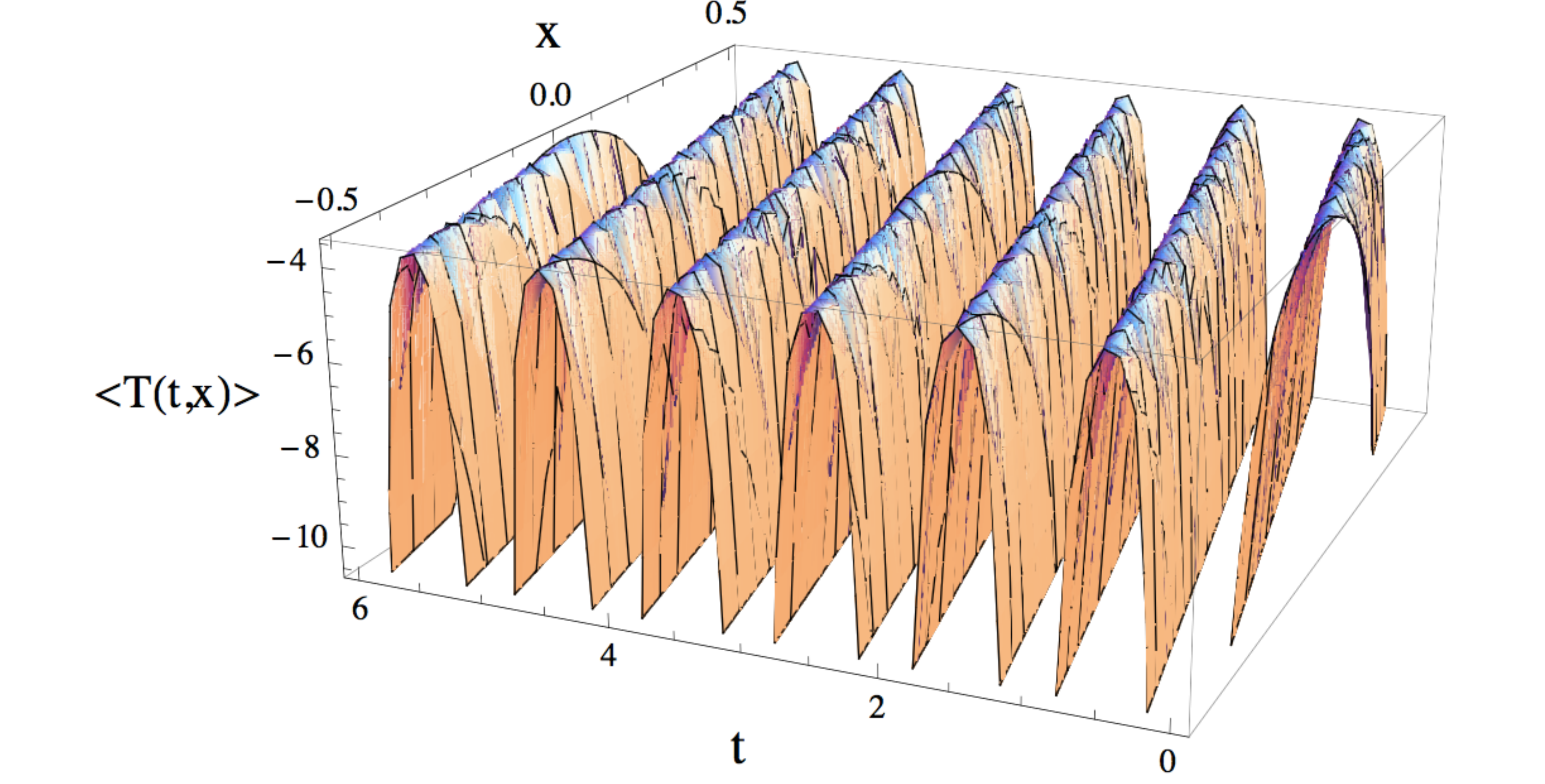}
\vspace{0.5in}
        \caption{Profile of the expectation value of the holomorphic stress tensor $T(t,x)$ across the rectangle.}
    \end{subfigure}
~
\qquad
    \begin{subfigure}[b]{0.35\textwidth}
        \includegraphics[width=\textwidth]{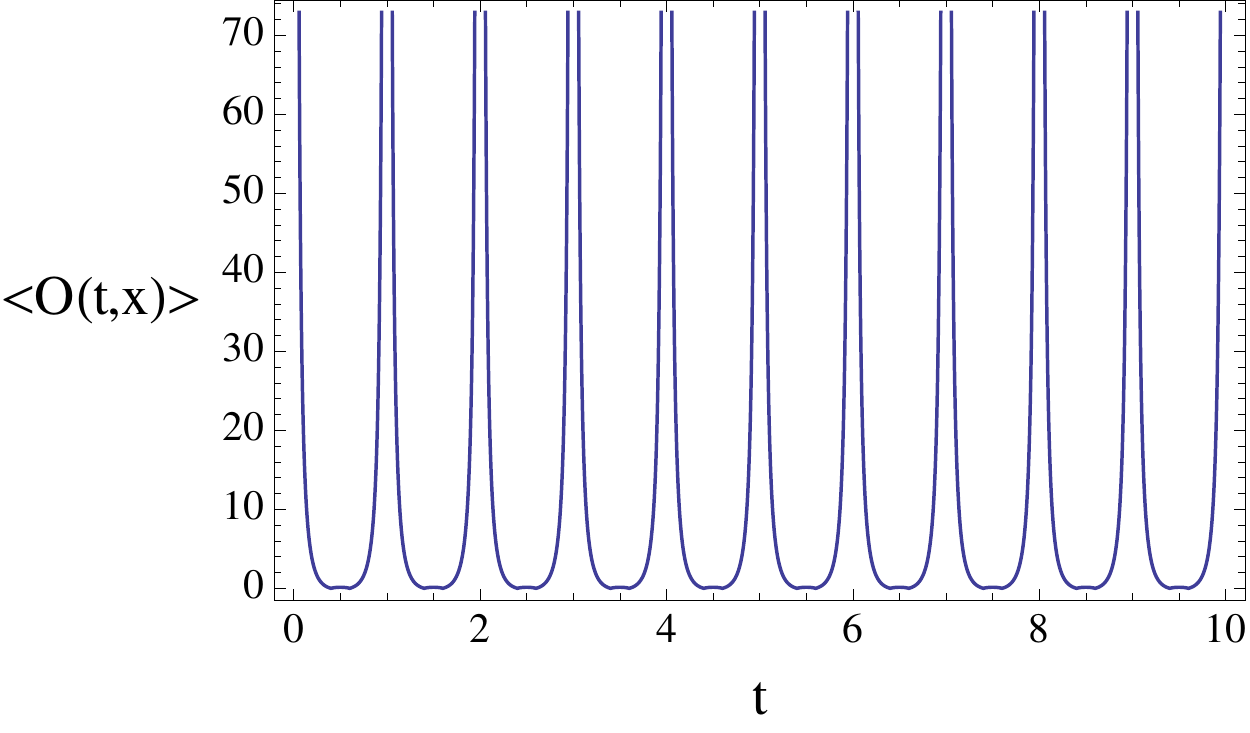}\\
        \includegraphics[width=\textwidth]{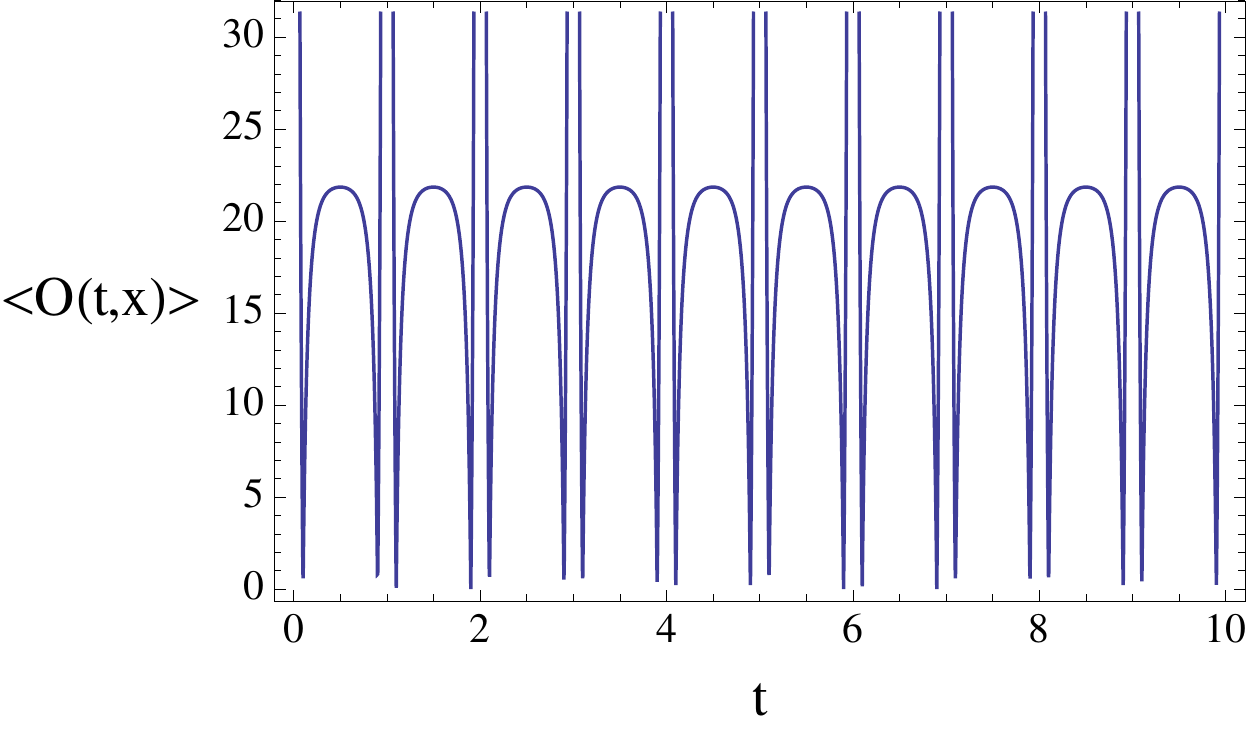}
        \caption{One-point functions $\left\langle \mathcal{O}(t,x)\right\rangle$ at sample points $x=0.1$ (top) and $x=0.4$ (bottom).}
    \end{subfigure}
    ~ 
    \caption{\small{Periodicity in observables in the rectangle ground state. Plot parameters are $k=0.5$ and $L=1$.}}
\label{fig:PeriodicObservables}
\end{figure}
Since correlation functions on the rectangle are calculated from their
counterparts on the UHP, every argument $z$ of a Lorentzian operator
assumes an inherent periodicity; e.g. one-point functions of primary
operators of conformal dimension $h$ in the conformal mapping of the UHP ground-state are given by
\begin{equation}
\left\langle \mathcal{O}(t,x)\right\rangle \sim\left.\left(\frac{dz(w)}{dw}\frac{d\bar{z}(\bar{w})}{d\bar{w}}\right)^{h}
\left(z(w)-\bar{z}(\bar{w})\right)^{-2h}\right|_{\begin{array}{c}
w\rightarrow x-t+iH/2\\
\bar{w}\rightarrow x+t-iH/2
\end{array}}
\label{eq:onePointMapped}
\end{equation}
 and the stress tensor is given by 
\begin{equation}
\left\langle T(t,x)\right\rangle =\frac{c}{12}\{z(w),w\},
\label{eq:stressTensorMapped}
\end{equation}
where $c$ is the central charge and $\{z(w),w\}=\frac{z^{(3)}(w)}{z'(w)}-\frac{3}{2}\left(\frac{z^{(2)}(w)}{z'(w)}\right)^{2}$
is the Schwarzian derivative. The periodicity of (\ref{eq:onePointMapped}) and (\ref{eq:stressTensorMapped}) in Lorentzian time
is therefore evident (Fig.~\ref{fig:PeriodicObservables}).
The periodicity of these observables as resulting from the nature of the conformal mapping and the implication that the
rectangle state features non-thermal behavior was also pointed out in \cite{Kuns:2014zka}. 

As we reviewed, conformal transformations of the vacuum
state on the upper-half plane do not thermalize, but one might
expect that perturbations of this boundary state should eliminate the non-thermal
behavior. Rectangle states perturbed by operator insertions in $c=1$ CFT were considered in \cite{Engelhardt:2015axa} in a somewhat different context. Below we investigate how the time evolution of similar states is affected by the spectrum of the CFT (rational vs. non-rational), and whether a given system may exhibit periodic revivals at asymptotically late times. Recall that, as was pointed out in Sec.~\ref{sec:thermprobes}, the expectation value of the stress tensor itself is in general an insufficient diagnostic of thermalization. In particular, in a state perturbed by Euclidean-time operator insertions $\mathcal{O}_i$ with conformal dimensions $h_i$, the time dependence in the expectation value of the stress tensor is determined purely by conformal invariance:
\[
\left\langle T(z)\prod_{i}\mathcal{O}_{i}(\zeta_{i})\right\rangle =\sum_{i}\left[\frac{h_{i}}{\left(z_{i}-\zeta_{i}\right)^{2}}+\frac{\partial_{\zeta_{i}}}{z_{i}-\zeta_{i}}\right]\left\langle \prod_{i}\mathcal{O}_{i}(\zeta_{i})\right\rangle, 
\]
since only the stress tensor coordinate $z$ is continued to Lorentzian time. As a result, the time evolution of this expectation value is qualitatively identical regardless of the spectrum of the CFT and cannot be used to resolve any potential differences for CFTs with $c<c_{\text{crit}}$ versus those with $c>c_{\text{crit}}$. In the next section we therefore probe perturbations of boundary states using one-point functions of generic operators that do not correspond to conserved currents, focusing on the different behaviors of rational vs. non-rational CFTs.

\section{Operator spectrum dependence of thermalization\label{sec:spectrumdependence}}

In this section we consider expectation values of primary operators in perturbed states. The simplest perturbed
states are those produced by a path integral over a suitable Euclidean domain with a single operator insertion on
the boundary of the domain. The expectation value of a single operator in such a state will then be given by
the analytic continuation of a three-point function with two operators on the boundary (one for the in-state and
one for the out-state) and one operator in the interior. Conformal invariance fixes the form of these
three-point functions up to a single unknown function of a suitable cross-ratio. Even without knowing the explicit
form of this function (which would involve knowledge of the structure constants and conformal blocks of the theory)
one can already see a qualitative change in the behavior of the part of the three-point function that is
determined by conformal invariance (and that we henceforth refer to as the ``universal'' part of the correlation function) as one moves from rational to non-rational theories. In particular, exact
periodicity of the expectation value appears to be lost\footnote{To see this, as we discuss below, 
we in fact need to consider linear superpositions
of states obtained by operator insertions.}, in agreement with the picture that rational theories
should not display global thermalization and irrational theories should. However, without more detailed knowledge
of the exact correlation function, it is not possible to see the destructive interference which leads to exponential
decay to the thermal value of one-point functions, and, while suggestive, our analysis is by 
no means to be taken as a proof of thermalization in irrational CFTs. 

\subsection{General setup}

It is in principle possible to consider very general classes of states created by a path integral over arbitrary bounded
domains with a particular boundary state on the boundary and arbitrary insertions of operators in the interior of the
domain and on its boundary. Even in the absence of operator insertions, correlation functions computed in states of this
type are in general time dependent. As we discussed in section~\ref{sec:thermprobes}, the time dependence in the expectation
value of the energy momentum tensor can in general be removed by applying a suitable diffeomorphism, and we will therefore focus
on the geometries with a time-independent expectation value for the energy-momentum tensor, which are infinite strip geometries.

We consider an infinite Euclidean strip of the form $w=x+i\tau$ with $(x,\tau)\in [0,2L]\times [-\infty,\infty]$, which can be mapped 
to the upper half plane via the map $z(w)=e^{\frac{\pi i w}{2L}}$, with $z$ the coordinate on the upper half plane. Such an
infinite strip can be interpreted in two different ways, either as providing an in- and an out-state on the theory on a
finite interval of length $2L$, but also as providing an in- and an out-state on an infinite spatial interval. In the latter
case, the roles of space and time should be exchanged\footnote{This is the strip state of CC~\cite{Calabrese:2006rx,Calabrese:2007rg}.}, so that Euclidean time runs from $0$ to $2L$ and space from $-\infty$ 
to $+\infty$. Moreover, the relevant analytic continuation to Lorentzian time is $w=x-t$ in the first case, and $w=L+it+i\tau$  
in the second case. We will mostly take the point of view of the finite strip in
what follows, but the infinite
strip can be treated in exactly the same way. 

We insert $n_1$ boundary operators $\mathcal{O}_B$ on the left boundary of the strip at $w_a=i\tau_a$,
and $n_2$ bulk operators $\mathcal{O}$ at positions $w_p=x_p+i\tau_p$. For simplicity, we will
not insert any operators on the right boundary of the strip, but this is a straightforward generalization. In order
to be able to interpret the boundary insertions as corresponding to an in- and an out- state, the boundary operators
should be distributed symmetrically around $\tau=0$. However, if we are interested in studying linear superpositions of
states, we should also consider asymmetric distributions of operators.

\subsection{Periodicity in correlation functions}

The general form of the correlation function can be obtained by mapping it to the upper-half plane and using $SL(2,\mathbb{R})$
Ward identities. To write the result we denote
\begin{eqnarray}
(\xi_1,\ldots,\xi_N)=(\{z_a(w_a)\},\{z_p(w_p)\},\{\bar{z}_p(\bar{w}_p)\}), \quad N=n_1+2n_2, 
\end{eqnarray}
in term of which the correlator is, up to an overall constant factor,
\begin{eqnarray}
\left\langle \prod_a {\cal O}_{B,a}(w_a) \prod_j {\cal O}_p(w_p) \right\rangle = \prod_{i}^{N} \xi_i^{h_i}
F\left( \frac{\xi_{ij} \xi_{kl} }{\xi_{ik} \xi_{jl} } \right) \prod_{i<j} \xi_{ij}^{\frac{2}{N-2}(h_{\Sigma}/(N-1)-h_i-h_j)}
\end{eqnarray}
where $h_{\Sigma}=\sum h_i$ and $\xi_{ij}=\xi_i-\xi_j$, and $F$ a function of cross ratios. Note that the conformal dimensions $h_i$ refer to both those of the bulk operator, $h$, $\bar{h}$ as well as those of the boundary operators, $h_B$. The prefactor 
$\prod \xi_i^{h_i}$ is due to the map from the strip to the upper-half plane and includes contributions from the coordinates of all operators\footnote{This prefactor is given by $\prod_i \left(\frac{dw_i(z_i)}{dz_i}\right)^{-h_i}$, which on the strip becomes $\prod_i \left(z_i(w_i)\right)^{h_i}$ up to a constant factor.}. It can be absorbed in a nice way
in the rest of the expression by defining
\begin{equation} \label{bb}
\tilde{\xi}_{ij} = \frac{\xi_i-\xi_j}{\sqrt{\xi_i \xi_j}} = 2 i \sin\left[ \frac{\pi}{4L}(w_i-w_j) \right],
\end{equation}
in terms of which the general correlator is of the form
\begin{eqnarray}
\left\langle \prod_a {\cal O}_{B,a}(w_a) \prod_p {\cal O}_p(w_p) \right\rangle = 
F\left( \frac{\tilde{\xi}_{ij} \tilde{\xi}_{kl} }{\tilde{\xi}_{ik} \tilde{\xi}_{jl} } \right)
\prod_{i<j} \tilde{\xi}_{ij}^{\frac{2}{N-2}(h_{\Sigma}/(N-1)-h_i-h_j)}.
\end{eqnarray}
Note that because of the the exponential map that we employ here has an explicit dependence on $i$ in it, 
\begin{eqnarray}
(w_1,\ldots,w_N)=(\{w_a\},\{w_p\},\{-\bar{w}_p\}).
\end{eqnarray}

Upon analytic continuation of a particular bulk operator to Lorentzian time, $w\rightarrow x-t$, it is clear from (\ref{bb})
that the correlation function will contain contributions of the form $f(t) =(\sin(\frac{\pi}{4L}(t-c)))^s$, with complex $c$,
which might appear to be periodic with period of at most $8L$, 
except that $s$ is in general not an integer and $f(t)$ has to be defined through
analytic continuation. For complex $c$, the function $z(t)=\sin(\frac{\pi}{4L}(t-c))$ follows a contour around the origin in
the complex plane that we can write as $z(t)=r(t) e^{i\phi(t)}$, with both $r(t)$ and $\phi(t)$ periodic with period $8L$.
The analytic continuation of $z(t)^s$ is clearly $r(t)^s e^{is\phi(t)}$, which is now no longer periodic unless $s$ is rational.
This is an indication that the time-dependence of correlation functions in rational theories will have special properties and
tend to be periodic. 

We will consider pure states of the form $\sum_i |\psi_i\rangle$ where each $|\psi_i\rangle$ is obtained
through a path integral on the half-strip with suitable operator insertions. Expectation values of bulk operators 
in such states require us to compute matrix elements $\langle \psi_i | \prod_k {\cal O}_k | \psi_j \rangle$.

We first focus on the diagonal matrix elements. For those, it turns out that the 
universal part of the correlation function will always be periodic.
To see this, we observe that the correlation function will contain a product of terms of the form
\begin{equation}
\left[\sin\left(\frac{\pi}{4L}(t-x+i\tau_{0})\right)\sin\left(\frac{\pi}{4L}(t-x-i\tau_{0})\right)\right]^{s},
\label{eq:pairing}
\end{equation}
which can be rewritten as the purely real expression 
\begin{equation}
2^{-s}\left[\cosh\left(\frac{\pi}{2L}\tau_{0}\right)-\cos\left(\frac{\pi}{2L}(t-x)\right)\right]^{s},
\end{equation}
which is well-defined with period $4L$.

Any possible breakdown of periodicity in diagonal matrix elements therefore 
has to originate from the function $F$ of the cross-ratios that
appears in the correlation function as well. Unfortunately, it is much more difficult to analyze this function in
general. If we take the simplest example with two boundary insertions at $\pm i\tau_0$ and one bulk operator, the cross-ratio 
(after analytic continuation) takes the form
\begin{equation}
y=\frac{\tilde{\xi}_{w,\bar{w}} \tilde{\xi}_{w_1,w_2} }{\tilde{\xi}_{w,w_1} \tilde{\xi}_{\bar{w},w_2}}=\frac{\sin\left[\frac{\pi}{2L}x\right]\sin\left[\frac{\pi i}{2L}\tau_{0}\right]}{\sin\left[\frac{\pi}{4L}(t-x-i\tau_{0})\right]\sin\left[\frac{\pi}{4L}(t+x+i\tau_{0})\right]}.
\label{eq:crossratio}
\end{equation}
We see that $y$ does not
go around one of the singularities at $y=0,1,\infty$ and that therefore the unknown function of the
cross ratio will remain periodic\footnote{We can also see this by observing that the denominator of (\ref{eq:crossratio}) can be expanded as $\frac{1}{2} \left(\left(a^2+b^2\right) \cos \left(\frac{\pi  x}{2 L}\right)-2 i a b \sin \left(\frac{\pi  x}{2 L}\right)-\cos \left(\frac{\pi  t}{2 L}\right)\right)$ where $a=\cosh \left(\frac{\pi  \tau_0}{4 L}\right)$ and $b=\sinh \left(\frac{\pi  \tau_0}{4 L}\right)$, so that it is given by the sum of a real time-periodic and a complex time-independent function.}  with period $4L$. Finally, for boundary operators inserted at $\pm\infty$ ($\tau_0\rightarrow\infty$), the cross ratio becomes time independent:
\begin{equation}
y = e^{\frac{\pi i x}{L}}-1,
\end{equation}
so that no decay in time can be seen for such operator placement. There may be an argument as to why diagonal matrix elements always remain
periodic based on the symmetry $\tau \leftrightarrow -\tau$, but we have not explored this in
detail.

Off-diagonal matrix elements, on the other hand, appear to lose their periodicity in general. This is already
clear at the level of the universal part of the correlation function, where factors of the form $[\sin
\frac{\pi}{4L}(t-c))]^s$, $c=x+i\tau_0$, are no longer paired with factors $[\sin
\frac{\pi}{4L}(t-c^*)]^s$ as in (\ref{eq:pairing}), resulting in an expression consisting of powers of periodic functions that are complex in the time argument. If $s$ is rational, these factors will remain periodic but with 
a longer period, but if $s$ is irrational periodicity is lost altogether. Of course, for a complete
analysis it is necessary to also consider the in general unknown functions of the cross-ratios\footnote{One
can easily check in examples of $c=1$ theories where correlation functions of primaries can be explicitly
written down that these conclusions indeed hold for the full correlation functions: for rational $c=1$ theories
periodicity is maintained, while for irrational $c=1$ theories periodicity is lost. (Note that periodicity is maintained if we take as our operators to be $\partial\phi$ or $\bar{\partial}\phi$, but since these operators correspond to conserved currents they should not be considered for diagnosing whether the system experiences revivals as previously explained.)  Because $c=1$ theories are
exactly solvable we do expect these theories to be described by a suitable Generalized Gibbs Ensemble
at late times under generic perturbation, see \cite{Mandal:2015kxi,Mandal:2015jla}.}. 
The analytic properties of correlation functions, and  more generally the analytic properties of
conformal blocks, are typically closely related to the braiding and fusion properties of the theory. 
For rational theories the space of conformal blocks form finite dimensional representations under fusion and braiding, 
which in turn is closely related to the periodicity of correlation functions\footnote{which we already knew to be periodic
in time anyway in view of the straightforward argument in the introduction.}. One would therefore expect to see
periodicity in the case of rational theories, and a breakdown of periodicity in irrational theories. As we have explained,
we already see signs of this breakdown in simple correlation functions, and it would be interesting to explore this further.

\section{The holographic dual of the generalized Gibbs ensemble}\label{sec:GGE}

As mentioned in the introduction, conformal field theories have a large number of conserved currents. 
For example, any polynomial made out of the 
stress tensor $T(z)$ and its derivatives is a conserved current. Similarly, if there are additional higher spin
currents, any polynomial involving those leads to conserved currents as well. Given such large sets of conserved
currents, one can ask what the maximal set of conserved and commuting charges is. For the case of the Virasoro algebra,
there exists a conserved current, unique up to total derivatives, whose zero modes all commute. In the semi-classical
case, where we replace OPEs by Poisson brackets, the construction of these conserved currents and corresponding
conserved charges is captured by the KdV hierarchy. The KdV hierarchy does in fact also describe the flows generated by
the complete set of commuting conserved charges. A conformal field theory contains a quantum deformation of the KdV 
hierarchy, the quantum KdV hierarchy, see \cite{Bazhanov:1994ft}.

Since the stress tensor is a single trace operator, adding polynomials of the stress tensor and its derivatives
with chemical potentials to the action (in order to describe a generalized Gibbs ensemble) 
corresponds to multitrace deformations in the CFT. Multitrace deformations
both in pure gravity as well as in its higher spin extensions can be conveniently studied in the Chern-Simons formulation, and
a detailed discussion will appear elsewhere \cite{deboerjottar}. Here we simply summarize a few key ingredients
using the notation from \cite{deBoer:2014fra}. 

In general, if we add a multitrace deformation of the form $\int \Omega\equiv \sum_i \nu_i F_i(W_s)$, with the $\nu_i$
chemical potentials which we will take to be constant, and $F(W_s)$ polynomials in the higher spin fields
and their derivatives, all we need in Chern-Simons theory is a boundary term of the form
\begin{equation}
I^{(E)}_B = -\frac{k_{cs}}{2\pi} \int_{\partial M} d^2 z\,{\rm Tr}\left[ (a_z+a_{\bar{z}})a_{\bar{z}} - 2\Omega\right]
\end{equation}
plus a similar result for the right movers. Moreover, whereas $a_{\bar{z}}$ usually contains the sources $\mu_s$ for the higher spin fields $W_s$,
we now need to replace these sources $\mu_s$ by $\partial\Omega/\partial W_s$. We therefore in general have a non-linear relation between
the normalizable and non-normalizable modes, which is typical for multitrace deformations \cite{Witten:2001ua,Sever:2002fk}.
The variation of the on-shell action consisting of standard Chern-Simons theory plus the boundary term can be written as
\begin{equation}
\delta( I^{(E)}_{CS}+I^{(E)}_B ) = \frac{k_{cs}}{\pi} \int_{\partial M} d^2 z\,{\rm Tr} \sum_i \delta\nu_i F_i(W_s),
\end{equation}
which indeed has the right structure.

Although we could continue our discussion in the language of Chern-Simons theory, from the above it should be
clear that the bulk field equations are not modified, and that once we restrict to translationally invariant solutions, 
in the bulk the solution looks just like the BTZ black
hole and its higher spin generalizations. This is also immediately the main point of this section: classically there are no
hairy black holes corresponding to the generalized Gibbs ensemble, the bulk geometry is still the BTZ geometry.
The free energy or partition function is however different 
from that of the usual BTZ black hole, because of the additional boundary terms that one needs. In fact, 
looking carefully at the Chern-Simons formulation, one finds that the contribution of the 
left-movers to the partition function for the pure gravity case with
a deformation 
\[
\int d\sigma \sum_i \mu_{i} F_i(T) 
\]
is equal to 
\begin{equation}
Z= {\rm Tr}(e^{-\int d\sigma \sum_i \mu_{i} F_i(T) }) = e^{2\pi \sqrt{\frac{c}{6} L_0} - \sum_i 2\pi \mu_i F_i(L_0)}
\left.\right|_{\rm saddle} .
\end{equation}
Here, saddle means that we have to extremize the right hand side with respect to $L_0$, and the answer therefore looks
like a generalized Legendre transform of the expression of the black hole entropy. 
Here, because we restrict to translationally invariant solutions, all terms containing derivatives of $T$ drop out
of $F_i(T)$, and these functionals become ordinary functions of the zero mode $L_0$. Thus, using the bulk gravitational
description simplifies the GGE dramatically, the zero modes of the higher spin conserved currents are polynomials in terms
of $L_0$ and no longer take on independent values. 

It is straightforward to see that 
\[
\frac{\partial \log Z}{\partial \mu_i} = - 2\pi F_i(L_0)|_{\rm saddle}.
\]
If we identify the $F_i(T)$ with the conserved charges of the KdV hierarchy, then the partition function is precisely
a tau-function for the KdV hierarchy\footnote{See, e.g. \cite{DJKM,JimboMiwa} for the tau-function and \cite{drinfel1985lie} for the KdV hierarchy.}.

Though this is perhaps a somewhat trivial example of a tau-function, we can extend it to the case where $T$ is no longer constant.
For this we need to use the gauge invariant generalization of the entropy given by the appropriate Wilson line operator.
The result is 
\begin{equation}
Z= {\rm Tr}(e^{-\int d\sigma \sum_i \mu_{i} F_i(T) }) = \exp\left\{\frac{c}{6} \cosh^{-1} \frac{1}{2} {\rm Tr}
P\exp\oint \left(\begin{array}{cc} 0 & 1\\ \frac{6T}{c} & 0 \end{array} \right) - \int d\sigma \sum_i \mu_{i} F_i(T) \right\}
\left.\right|_{\rm saddle} 
\end{equation}
where now saddle means that we should find the saddle point of the expression on the right hand side for the 
functional $T(\sigma)$. This provides a more interesting class of tau-functions for the KdV hierarchy if the
$F_i(T)$ are the corresponding charges, but at this level the $F_i(T)$ can in principle still be arbitrary,
which is probably an artifact of the large $c$ (or large $k$) limit.
We expect that once we start quantizing Chern-Simons theory with matter we should see a more interesting structure emerge, 
and, in particular, we expect gravitational solutions that depend non-trivially on the chemical potentials $\mu_i$.
It would be interesting to explore the construction of such ``black holes with quantum hair'' in more detail.

Finally, we note that while it is tempting to assume a connection between the conserved charges considered here 
and the conserved charges that appear in studies of integrability in AdS/CFT, the latter are generically non-local
and are supposed to already be relevant at the semi-classical level. Therefore, an obvious connection is lacking, but
it would also be interesting to explore this in more detail, as would be the role that the various conserved charges can 
possibly play in studying geon solutions and instabilities of AdS.

\section{Discussion\label{sec:discussion}}

In this paper we studied some properties of the non-equilibrium behavior of 2D CFTs as well as the distinction
between local and global thermalization. We provided arguments that there are no revivals in generic states in
irrational theories, which one could take as an indication that the system thermalizes. To actually see thermalization
probably requires one to choose very complicated initial states for which explicit computations rapidly
become intractable. One-point functions of light probes in very complicated, heavy states can presumably 
be well-approximated by the light-light-heavy-heavy conformal block derived in \cite{Fitzpatrick:2015zha},
although these computations have to our knowledge not been extended to a situation with boundaries.
Ultimately this is just another illustration of the usual problem than we can either do explicit, weakly coupled computations
where unitarity is manifest but thermalization difficult to see, or we can do strongly coupled (e.g. gravitational)
computations where thermality is easy to see but manifest unitarity is lost. 

We note that there has been much research in recent years, starting with \cite{Bizon:2011gg,Dias:2011ss,Dias:2012tq,Maliborski:2013jca}, into the possibility of time-periodic solutions in AdS that avoid collapse into a black hole; however, exact solutions involving stable oscillating matter known to exceed the BTZ black hole mass threshold (in AdS$_3$) and yet exhibiting revivals to $t\rightarrow\infty$ (undamped oscillations) have so far not been found. If such solutions do exist, they appear likely to occupy a very small phase space and/or involve considerable simplifications of the physical setup.

In a similar spirit, we have shown that the holographic dual of the
generalized Gibbs ensemble is still a BTZ black hole. The GGE has
been central to the discussion of 1+1-dimensional integrable systems
away from a conformal fixed point. Such integrable field theories
can be obtained as massive deformations of a CFT, and -- at least
in principle -- the analysis that was carried out here could be applied
to them via conformal perturbation theory, where transformations to
a frame of constant stress tensor can still be applied at every order.

We note several additional avenues that are of interest in light of
our findings. The holographic picture of Sec.~\ref{sec:thermprobes} for
diffeomorphisms of the CFT ground state can be used to generalize
the AdS/BCFT setup of \cite{Takayanagi:2011zk,Fujita:2011fp} to arbitrary
forms of boundary states by finding the appropriate bulk brane corresponding
to the extension into the bulk of the dual BCFT's boundary. In particular,
the holographic dual of the rectangle state can thus be found, and
Lorentzian-time correlators from the corresponding initial state can
be computed via a formalism such as \cite{Skenderis:2008dh,Skenderis:2008dg}.
The holographic implementation of such a setup would likely be a useful
tool in evaluating general non-equilibrium behavior in systems with boundaries,
not only in classical AdS geometries, but also to $1/N$ corrections. 
Finally, we note that while in classical $SL(2,\mathbb R)$ Chern-Simons theory
expectation values of Wilson lines in different representations are related to each
other in a simple way, this is no longer the case in quantum Chern-Simons theory. 
It would be interesting to explore these quantum expectation values in more detail
and establish their relationship to the quantum KdV hierarchy and the GGE at finite
values of the central charge.

\subsubsection*{Acknowledgments}
We are grateful to G. Barnich, N. Engelhardt , B. Freivogel, D. Harlow, T. Hartman, J. Jottar, and D. Marolf for useful discussions. 
DE acknowledges support from NSF grants PHY-1313986 and DGE-0707424, the Netherlands Organisation for Scientific Research (NWO), and the University of Amsterdam and is grateful to the Kavli Institute for Theoretical Physics for hospitality during the workshop Quantum Gravity Foundations:  UV to IR while this work was in progress.

\bibliographystyle{utphys}
\bibliography{thermCFTrefs}

\end{document}